# Automated Solar Feature Detection for Space Weather Applications

**David Pérez-Suárez, Paul A. Higgins, D. Shaun Bloomfield, R.T. James McAteer, Larisza D. Krista, Jason P. Byrne and Peter. T. Gallagher.**
*School of Physics, Trinity College Dublin, Dublin 2, Ireland*

## ABSTRACT

The solar surface and atmosphere are highly dynamic plasma environments, which evolve over a wide range of temporal and spatial scales. Large-scale eruptions, such as coronal mass ejections, can be accelerated to millions of kilometers per hour in a matter of minutes, making their automated detection and characterisation challenging. Additionally, there are numerous faint solar features, such as coronal holes and coronal dimmings, which are important for space weather monitoring and forecasting, but their low intensity and sometimes transient nature makes them problematic to detect using traditional image processing techniques. These difficulties are compounded by advances in ground- and space- based instrumentation, which have increased the volume of data that solar physicists are confronted with on a minute-by-minute basis; NASA's *Solar Dynamics Observatory* for example is returning many thousands of images per hour (~1.5 TB/day). This chapter reviews recent advances in the application of images processing techniques to the automated detection of active regions, coronal holes, filaments, CMEs, and coronal dimmings for the purposes of space weather monitoring and prediction.

## INTRODUCTION

Astrophysics seeks to determine the physical properties of celestial bodies, primarily by studying the light they emit. This is achieved using remote observations as the distances are generally too great to allow in-situ measurements. Our Sun is the closest of all stars, by many orders of magnitude, and allows scientists to perform long-baseline synoptic studies at size scales which are impossible with other stellar objects. The Sun is also the source of life on Earth, making the study of the Sun-Earth interaction extremely important, especially in our technology-dependent society of today. The study of space weather focuses on disturbances produced by the Sun and the effects that they have on the environment near Earth, the other planets, and throughout the heliosphere. Those disturbances can affect satellites, airplane communications, long metallic oil pipe lines and electrical distribution grids, to name a few. More directly, it can affect the health of air crews and passengers on polar flights and astronauts. Accurate forecasting of those disturbances and their effects allows us to prepare for their arrival. The Sun is routinely observed by numerous ground- and space- based observatories and the study of features in real time (i.e., those currently on the solar surface) provides a better insight into what may happen at a later time elsewhere in the heliosphere.

When Galileo Galilei turned his telescope to look at the Sun in the early 17th century he became one of the first scientists to look in detail at the solar atmosphere. He pointed out that sunspots are features on the surface of the Sun and used them to study solar rotation. Since then, many observatories have studied, counted, and classified sunspots as they emerge and evolve on the Sun. These observations have been used to understand more than how the Sun rotates; historical data have made important contributions in studying the 11-year solar activity cycle and even some possibly-related Earth climate changes (e.g., the "little ice age" in Europe during the latter half of the 17th century occurred in the Maunder minimum, an almost 60-year period in which the Sun seemingly produced few sunspots, (Eddy 1976; Lockwood et al. 2010). Images were drawn by hand and classified by eye in those early observations, initially using pencil drawings before photographic plates became common use. The invention in 1969 of the Charge-Coupled Device (for which Willard S. Boyle and George E. Smith won the 2009 Nobel prize in physics) was the start of a new era for solar physics. The ability to directly digitize images at their acquisition allowed telescopes to rapidly acquire data. Shortly thereafter, space-based missions started taking observations in different wavelengths, providing a more complete view of the Sun. The many different types of instruments on-board spacecraft (i.e., imaging, spectrograph, and in-situ detectors) have also provided invaluable information for a recent field of study called "space weather". Space weather generally refers to the combined effect that all forms of solar activity have on objects within the heliosphere (including planets, their atmospheres, and satellites.) The ultimate aim of space weather research is to accurately forecast the arrival time of events which affect the heliosphere. To achieve this we need a better understanding of the different features that appear on the Sun and the resulting different forms of solar activity.

In general the solar atmosphere is stratified into temperature layers, each of which can be distinguished by the dominant type of radiative emission. The coldest, and lowest, layer emits mostly visible light (i.e., the photosphere, approximately 6,000 K) while the hottest, and highest, layer emits mostly in extreme ultraviolet (EUV) and X-ray wavelengths (i.e., the corona, more than 1 MK; Stix, 2004). Fortunately for human life, the Earth's upper atmosphere blocks most of the high energy solar radiation. However, this makes it impossible to observe the hottest layers from ground-based facilities. The observation of these layers is achieved with the use of instruments on-board high-altitude balloons, rockets, or spacecraft. A variety of instruments are used to study the processes which occur on the Sun. Imaging devices are generally sensitive to restricted wavelength range, while spectrometers provide information across wavelength at the expense of losing a spatial dimension. Another important instrument is the magnetograph that measures the magnetic field (or a component of this, e.g., the longitudinal component along the line-of-sight) at some height within the solar atmosphere (typically at the photospheric surface). Finally, coronagraphs are used to study the immediate environment around the Sun. These instruments consist of an imaging unit with an occultation disk that obscures the solar disk. The reader is referred to Stix (2004) as well as the relevant documentation for each instrument (specific papers are given in the following chapter sections) for a more detailed description of solar instruments and their characteristics.

The following sections provide an overview of the main feature detection techniques that are used for space weather forecasting. The features are classified into two groups depending on the form of data necessary to detect them: spatial features observed in single images (e.g., sunspots); temporal features requiring more than one image to characterise them (e.g., transient events, such as coronal mass ejections). However there is an overlap, as some of the techniques used to segment the first kind of feature do require more than one image in order to provide a robust threshold. The identification of many other features on the Sun, not related to space weather and hence not discussed shown in this chapter, may be found in the literature reviews of image processing techniques applied to solar images. Sanchez et al. (1992) provides a substantial review on the techniques used mainly for ground-based observations, e.g., to reduce the effect of the atmospheric turbulence around the telescope. High-resolution, ground-based images have been used to detect and track the granular (and supergranular) cells on the photosphere (e.g., Rieutord et al., 2007; Potts et al., 2004). Zharkova et al. (2005) give a very detailed description of some of the features detected as part of the European Grid of Solar Observation (EGSO) programme, while Aschwanden (2010) provides a general review of techniques used to detect a multitude of solar features.

## DETECTION OF SPATIAL FEATURES

This section focuses on the detection of time-independent features. This does not preclude evolution of these features; in fact they must evolve as there is nothing truly static on the Sun. However, they may be detected without knowledge of how they evolve. The dynamics of these features can then be measured by following them across the solar disk as the sun rotates. The description of a few detection codes focusing on active regions, coronal holes, and filaments are discussed.

### Active regions

Solar active regions were historically observed as sunspot groups in the photospheric continuum. Physically, active regions are concentrations of magnetic flux that have emerged through the solar surface and are thus a manifestation of solar magnetic activity. Sunspots were the first solar feature to be cataloged and have been routinely measured by multiple observatories over the last four centuries. Observations obtained during the last 60 years have highlighted the different forms which active regions appear to take when observed in different wavelength ranges (i.e., at different temperatures). In 1972 the National Oceanic and Atmospheric Administration (NOAA[1]) started a sequential numbering system for active regions. Figure 1 shows an example of NOAA 10708 extracted from SolarMonitor.org (Gallagher et al., 2002). It is clear from the figure that different algorithms must be used to identify active regions at each height in the solar atmosphere. Sunspots are the signatures of active regions in the visible photosphere (Fig. 1a), normally with a dark inner umbra and a surrounding penumbra. In the warmer overlying chromosphere they are observed as extended bright patches in visible and UV emission (Figs. 1c and 1d). Meanwhile, in the hot higher corona they appear as high-contrast regions of EUV emission where loop-like structures can often be distinguished. The varied appearance at different temperatures (and thus heights) in the solar atmosphere makes a consistent definition of an "active region" a somewhat difficult, and occasionally controversial, task. It should be noted that NOAA only designates numbers to those regions which have a white-light signature (i.e., a sunspot in the visible continuum)[2].

As is evident from Figure 1, the same detection technique cannot be used for every image across all wavelengths. However, for cataloguing purposes there is no need to detect the feature in every possible wavelength. When multiple channels are used as an input the segmentation can be done through supervised or unsupervised classification. The main difference is in how the classes are selected: the supervised method requires training values whereas classification values in the unsupervised approach are determined by the algorithm. From a space weather perspective, images of the magnetic field (i.e., magnetograms), provide the most information. The detection of active regions is essential because they are the source of several forms of solar activity, such as flares and coronal mass ejections (CMEs).

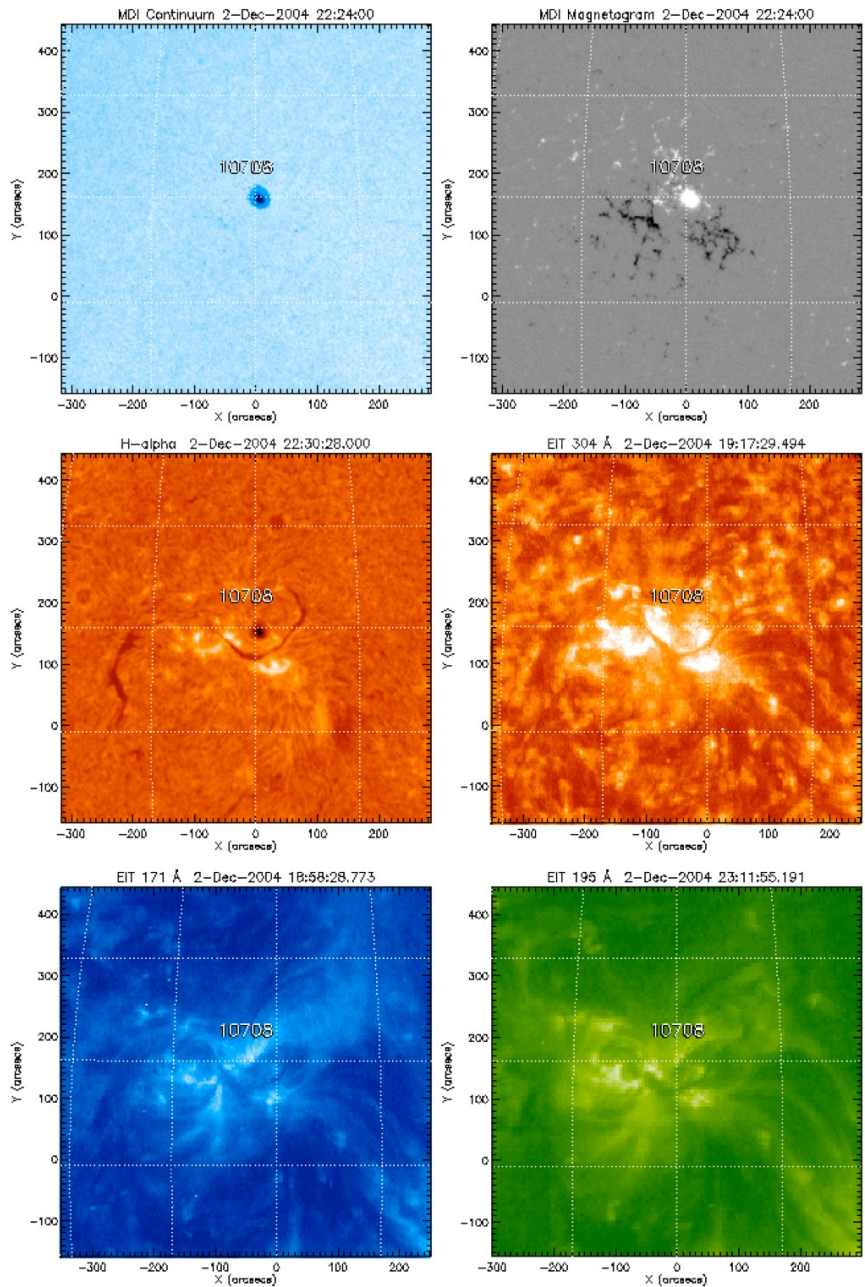

***Figure 1.*** *These six images show NOAA active region 10708 as viewed in (a) continuum (photosphere, ~6,000 K) from MDI, (b) magnetogram from MDI, (c) H-alpha (chromosphere, ~8,000 K) from Big Bear Solar Observatory and d) 304 Å (higher chromosphere, ~10,000 K), e) 171 Å (corona ,1 MK) and f) 195 Å (corona, 1.2 MK) from the Solar and Heliospheric Observatory. Extracted from http://www.solarmonitor.org (Gallagher et al., 2002).*

Currently, region-based flare and CME forecasting requires determination of the magnetic properties of an active region in question and its surroundings (e.g., Conlon et al., 2008; Zhang et al., 2009 and references therein). Turmon et al. (2002, 2010) describe a Bayesian technique to segment active regions in both ground- and space- based magnetogram data. The SolarMonitor Active Region Tracker (SMART, Higgins et al., 2010) is an algorithm for detecting, tracking, and cataloging active regions throughout their emergence, evolution, and subsequent decay. It extracts magnetic properties such as active region size, total magnetic flux, flux imbalance, growth or decay rate, and measurements of magnetic morphology. The SMART code operates in four main steps. First, the magnetograms are segmented into individual feature masks (Figure 2). Second, a characterization algorithm is run on each extracted region to determine its physical properties. Third, extracted regions are classified using a simple scheme. Finally, the regions are catalogued and tracked through time. Here we are interested only in the initial segmentation technique, so the reader is referred to the SMART documentation (Higgins et al., 2010) for further information on the other aspects of the algorithm.

SMART uses two consecutive magnetograms to allow for the removal of transient features as well as the extraction of time-dependent properties. The first steps applied, shown in Figure 2, are smoothing of the images (top row, middle column) with a 2D Gaussian and removal of the background using a static threshold (top row, right column). Binary masks are created from the corrected magnetograms (second and third rows, left column), setting all pixels above the threshold to one. The masks are radially dilated (second and third rows, middle column) and subtracted in order to identify and remove transient features (right column, middle row). The resulting mask is then radially dilated (bottom row, middle column). Each feature (bottom row, right column) is then characterised individually by its physical properties, which are determined from the later magnetogram.

The SMART algorithm is unique among automated active region extraction algorithms (e.g., McAteer et al. 2005a), in that it facilitates the temporal analysis of magnetic properties from first emergence of an active region through tracking the reappearance of regions over multiple solar rotations. In contrast, NOAA assign new numbers to active regions that rotate around the east limb onto the visible solar disk, irrespective of whether they are newly emerged or previously existing active regions.

Future revisions of SMART will incorporate a flare event probability. This will be determined using active region properties determined by SMART, including a measure of magnetic flux near polarity separation lines (Schrijver, 2007) and a proxy for non-potentiality (Falconer et al., 2008). The idea of using many active region properties to determine flare event probabilities was explored in Leka & Barnes (2003). The accuracy of flare prediction may increase by adapting more esoteric AR properties, such as the McAteer et al. (2005b) fractal and Conlon et al. (2009, 2010) multifractal techniques to the SMART feature characterisation. Conlon et al. (2010) propose a 2D wavelet transform modulus maxima method to study the multifractal properties of active region magnetic fields whereby the segmentation of the region is provided by an adaptive space-scale partition of the fractal distribution that shows a potential link to the onset of solar flares.

Other algorithms seek to provide automated classification of sunspot groups in the same framework as human observers. Colak & Qahwaji (2008) use white-light images and magnetograms from the Michelson Doppler Imager (MDI; Scherrer et al. 1995) onboard the *Solar and Heliospheric Observatory* (*SoHO*; Domingo et al., 1995) with neural network techniques to detect sunspots and classify active regions according to the McIntosh classification system. This system has the advantage that it achieves similar results to the NOAA active region identification scheme.

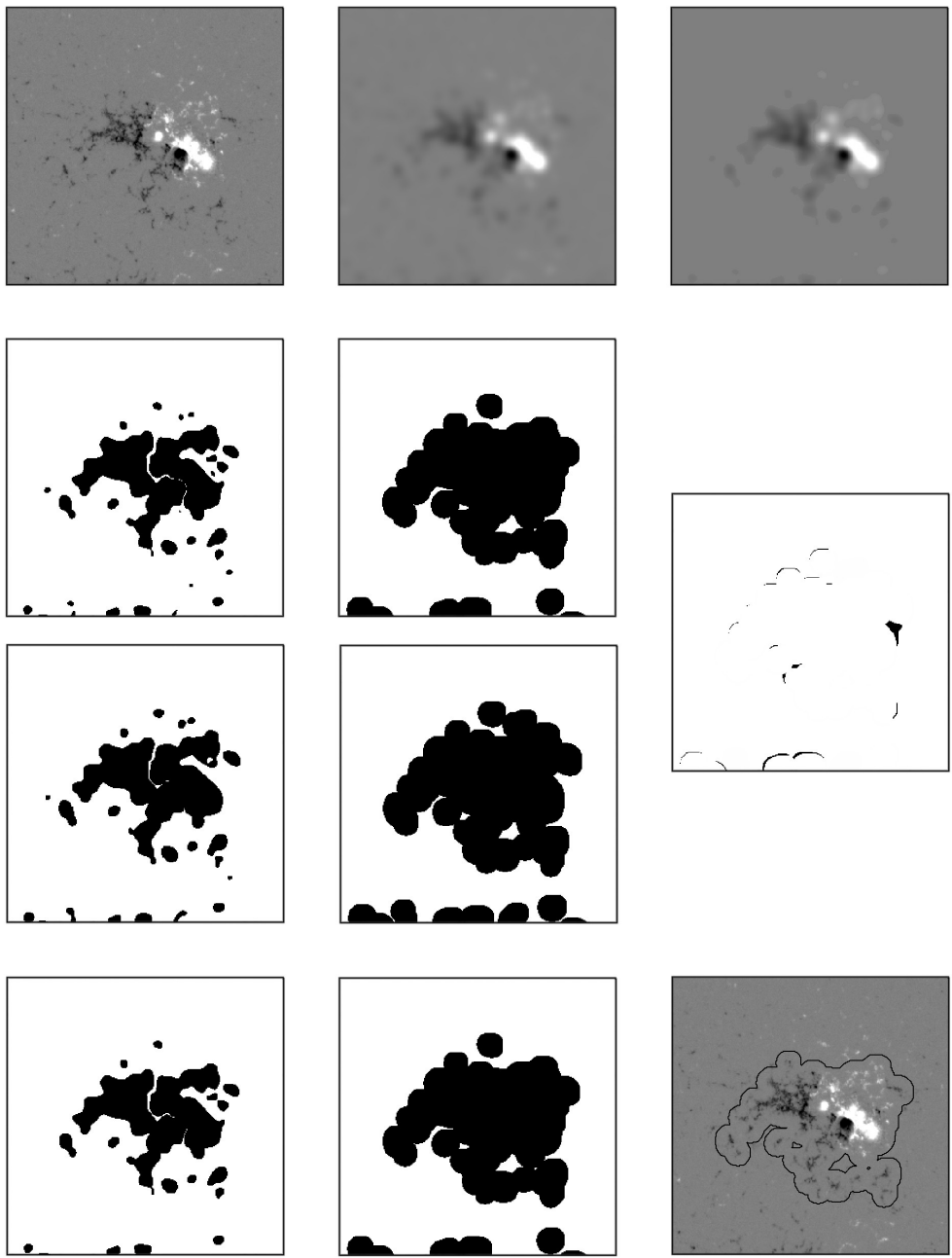

*Figure 2.* SMART steps. First image, top-left, shows the magnetogram of an active region. Next images on the right show the smoothing and thresholding at 70 G respectively. Next two rows show the masks from that image and the one obtained 96 minutes before, the growing step, and their difference. The last row shows the first mask after subtracting the difference and after dilation, which is the final segmentation mask used to plot the contours on the last image.

Zharkova et al. (2005) discuss two codes (developed for EGSO at Meudon observatory) to study active regions. Information about sunspot properties (i.e., size of umbra and penumbra) and intrinsic magnetic properties are obtained by comparing ground-based images of chromospheric emission (i.e., Ca II K1) and *SoHO*/MDI white-light images with *SoHO*/MDI magnetograms. These two codes are complementary: one extracts the magnetic field properties of the active region; one extracts the properties of the sunspots. The segmentation is obtained using a *Sobel* edge detection technique on the photospheric images. A global threshold segments the edges and the existing gaps are filled with the *close* and *watershed* morphological operators, followed by a new segmentation based on dynamic thresholding (constant for the MDI data; variable for ground-based images, due to Earth's unstable atmospheric conditions) to extract the sunspot umbra and penumbra. Not all active regions detected by SMART produce a photospheric sunspot signature in the continuum. Therefore, studying the output of both SMART and the Zharkova codes will help in the understanding of the production and evolution of the solar magnetic field.

Active regions can also be extracted from a number of EUV passbands. Dudok de Wit (2006) discuss a supervised clustering method for which some applications are shown in Chapter **??**. Barra et al. (2009) use a fuzzy clustering technique called Spatial Possibilistic Clustering Algorithm (SPoCA). SPoCA is a multichannel, unsupervised, spatially-constrained, fuzzy clustering method that automatically segments solar EUV images into regions of interest. It has the ability to detect multiple features at once, such as active regions, coronal holes, and quiet Sun. The nature of this code allows the detection of additional features, such as filaments and coronal bright points, and the addition of images observed in other wavelengths. The algorithm attempts to find the cluster centres of the features being detected through an iterative minimization equation. Each pixel obtains a probabilistic value of belonging to one or other feature group which depends on itself, its closest neighbours, and the whole image. SPoCA includes a radial line-of-sight equalization, inclusion of an automatic evaluation of the segmentation with a *sursegmentation* method (i.e., segmenting the image into a number of classes strictly superior to the intuitively expected number of classes in the image, and then finding an aggregation criterion of the resulting partition that shows the relevant classes), and smoothing of the edges using a morphological opening with a circular isotropic element.

**Coronal holes**

Coronal holes are low-density regions in the hot, high-lying solar corona that exhibit reduced EUV and X-ray emission when compared to the quiet Sun and active regions. The magnetic field distribution within coronal holes is believed to be dominated by a single polarity. This is probably due to the predominantly open nature of the magnetic field lines that extend beyond the corona and into the interplanetary medium. As a result, coronal holes give rise to the high-speed solar wind streams (Altschuler et al., 1972), causing recurring magnetic disturbances at Earth on time scales of days to months as these streams sweep past. Coronal holes can be observed in EUV and X-ray wavelengths from rocket or space-based telescopes as well as in the cooler, lower-lying chromospheric He I 10830 Å infrared absorption line from ground-based telescopes. They appear dark in EUV and X-ray because of a low emission-line strength (caused by reduced densities), while they are bright in He I because of a low absorption-line strength (caused by reduced population of the atomic state required for radiative absorption). In order to automate coronal hole detection, various teams have developed approaches mostly based on threshold segmentation. Two detection algorithms are described below: the first detects CHs from Earth using ground-based He I images, while the second uses EUV images from the Extreme ultraviolet Imaging Telescope (EIT; Delaboudinière et al., 1995) onboard *SoHO*.

Ground-based observations from the Kitt Peak Vacuum Telescope (KPVT) permitted the cataloging of coronal holes from 1974 to 2003 through manual identification. Henney & Harvey (2005) developed an algorithm motivated by the conclusion of operations of the KPVT in 2003 and the start of the synoptic observations by the SOLIS Vector Spectro-Magnetograph helium spectroheliograms and photospheric magnetograms. The method uses a two-day averaged He I 10830 Å spectroheliogram and a two-day averaged photospheric magnetogram, weighted by an expression involving their time difference. A mask value is determined as the 10% level of the median of positive values. A morphological *closing* operation (using a square kernel function as the shape operator) is applied to fill the gaps and connect nearby regions. For physical reasons, explained in Harvey & Recely (2002), areas smaller than two supergranules are removed and the mask is then multiplied by a very large value to fill the corresponding pixels on the first segmented image. The image is then smoothed to fill in small gaps and holes. This is followed by the *open* morphological operation which removes small features while preserving the size and shape of the detected regions. The magnetic properties of each candidate region are extracted from the magnetograms and the percentage of unipolarity is examined in order to disregard those with a value below a varying threshold.

More recently, Krista & Gallagher (2009) developed a coronal hole identification method that compares coronal hole properties with in-situ solar wind properties at ~1 AU. The algorithm also incorporates a space weather forecasting tool to predict the arrival of the fast solar wind streams at Earth using the Parker solar wind model (Parker, 1958). The coronal hole boundaries achieved are similar in EUV and X-ray wavelengths and agree well with the boundaries determined by eye. The automated high-speed solar wind forecasts are also in good agreement with the observed high-speed solar wind arrival times determined from in-situ solar wind measurements. In this method, intensity histograms of a solar EUV image give a multimodal distribution, where each frequency distribution corresponds to a different form of feature on the Sun - i.e., low intensity regions, quiet Sun, and active regions. The intensity of the coronal hole boundary corresponds to the location of a local minimum between the low intensity region and quiet-Sun distributions. As Figure 3 shows, this local minimum can be enhanced using a partitioning operation (1st and 2nd row in Fig. 3). Through such an approach, the local histograms obtained for each sub-image have more defined minima, which aids in determining the global threshold. This method works for any time in the solar cycle regardless of the change in the overall solar intensity, as it depends solely on the mean quiet-Sun intensity within the image in question. However, the coronal hole boundaries acquired during solar maxima may be less accurate due to bright coronal loops intercepting the line-of-sight and obscuring parts of the coronal hole boundaries.

After segmentation, low intensity regions are classed as either coronal holes or other dark quiet-Sun features (e.g., filaments) using magnetogram data; filaments have a balanced bi-polar distribution (i.e., close to zero skewness) of magnetic flux, whereas coronal holes have a dominant polarity (i.e., an imbalanced bi-polar distribution with a relatively large skewness). The physical properties of each coronal hole are automatically determined for forecasting purposes. For each coronal hole group the arrival time of the corresponding high-speed solar stream is determined at Earth and the predicted and observed arrival time is then monitored for further development of the method.

The results of the two algorithms described above have not yet been compared, however both obtain satisfying results when compared with other sources. Henney & Harvey (2005) compared their results with the hand-drawn coronal hole maps and found area differences of 3% or smaller. Krista & Gallagher (2009) have compared their high-speed solar wind arrival times with observed arrival times and obtained a positive correlation between the high-speed solar wind duration and the coronal hole area.

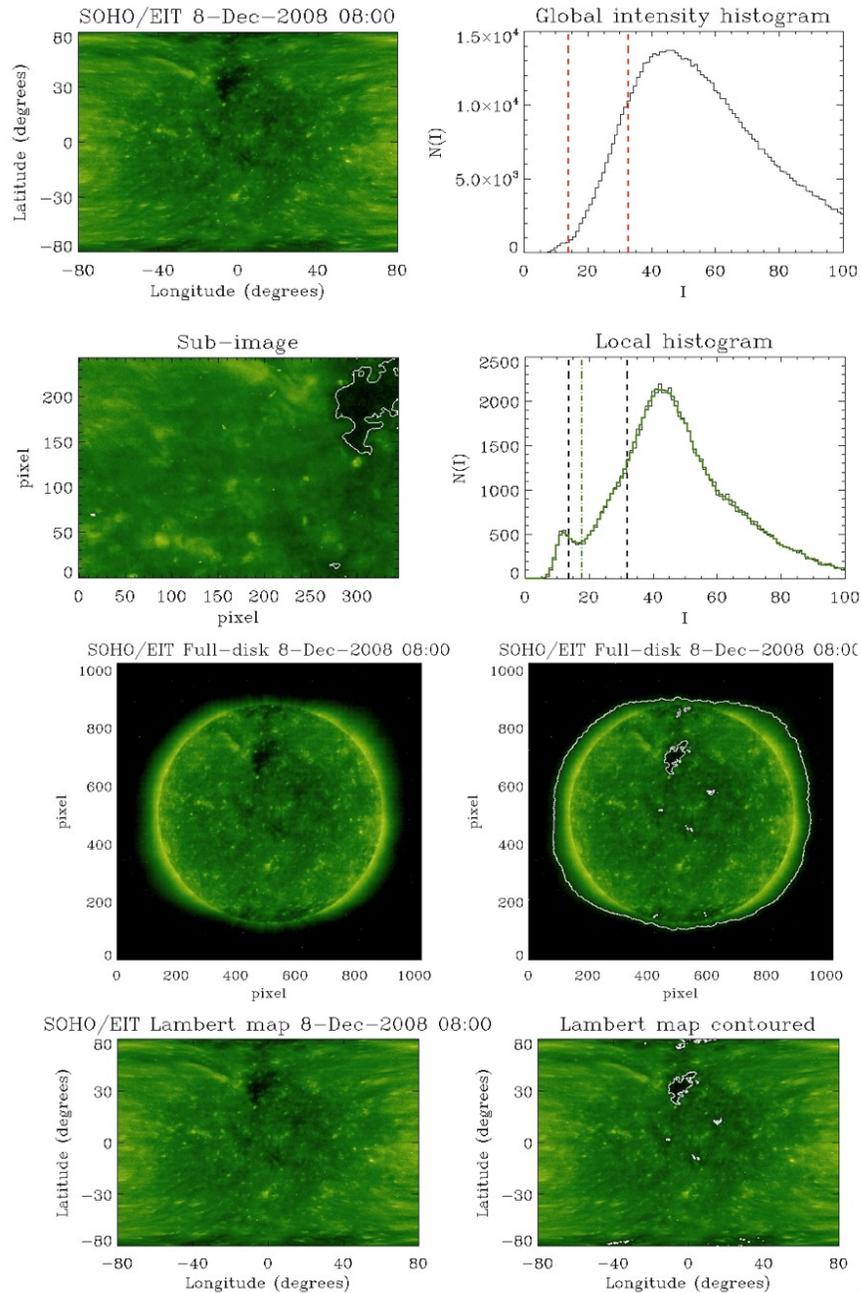

*Figure 3.* *Visualization of the coronal hole detection algorithm developed by Krista & Gallagher (2009). The 195 Å SoHO/EIT full-disk image is transformed to a Lambert equal area projection map (1st row, left). The corresponding global intensity histogram is obtained (1st row, right), as it has a unimodal distribution, no coronal hole threshold can be obtained. The map is then divided into sub-images (2nd row, left) to obtain local intensity histograms (2nd row, right) which are more likely to have a bimodal distribution. In all histograms the black and red dashed lines give the range where the threshold is searched for, and the dashed green line is the threshold found. The 3rd and 4th row images show the SoHO/EIT 195 Å full-disk image and Lambert projection map respectively with and without the low intensity region contours.*

# Filaments

Filaments are large volumes of very dense, cool plasma held in place by magnetic fields. They usually appear as long, dark, and thin features when observed against the solar disk, whereas they appear as bright, fuzzy arches and are called prominences at the limb. Images at chromospheric temperatures (particularly in the optical H-alpha line) provide the best outline of these features, even though filaments are also observed in the corona. H-alpha observations are routinely made from ground-based telescopes (the recent *Hinode* spacecraft does contain an H-alpha filter but, due to problems arising during launch, its use is not recommended). The images thus require pre-processing to correct for the constantly varying observing conditions caused by atmospheric seeing. Some of the corrections performed are the same for all ground-based observations, but others are instrument dependent. The reader is referred to the documentation of each algorithm for further details.

It is well established that the sudden disappearance (or eruption) of a filaments is usually associated with a CME. The characterisation of filaments can provide information with which to predict the orientation of the magnetic field associated with CMEs and hence the probability of a CME being geo-effective (i.e., its likelihood for impacting Earth). Bernasconi et al. (2005) produced a very complete, automated filament detection and characterization algorithm that is based on an existing code by Shih & Kowalski (2003). Their approach uses full-disk H-alpha images observed from Big Bear Solar Observatory (BBSO), such as the one shown in the left panel of Figure 4. The filament detection is performed by creating a mask using both threshold segmentation and an advanced morphological filtering operation. The first step removes the sunspots (their cores, or umbrae, are usually darker than filaments). This requires a filtering operation to extract only those regions with elongated shapes. These shapes are isolated by separately applying eight opening morphological operations to a filament mask with the eight linear structuring elements shown in the top-right panel of Figure 4 (Soille & Talbot, 2001). Pixels that survive at least two of these opening operations are used as seeds for a region-growing morphological filter and regions smaller than 300 pixels are deprecated. Once the mask has been created, each separate cluster is numbered and the characterisation of the detection proceeds.

In this process the position, length, area, average tilt of axis and chirality of the magnetic flux rope are extracted. The determination of the filament spine is performed using a multi-step iterative technique, shown in the bottom-right panel of Figure 4. The first iteration starts by determining the location of the two spine end points. Then it determines another vertex by adding the middle point and applying an optimization process. These steps are iterated, resulting in an array with the coordinates of the filament's spine.

The barbs of a filament are an important characteristic to take into account because they may yield information on the chirality of the flux rope within which the filament is embedded. The angle of each barb relative to the closest spine segment determines them as bear-left or bear-right. The difference between the number of bear-left and bear-right barbs establishes the chirality of the filament as left- or right- handed.

Other automatic filament detection techniques differ mainly on how the threshold is selected. In the EGSO algorithm (Fuller et al., 2005) the lower and upper thresholds used for finding seeds to create a mask in the segmentation of the image are calculated according to local statistics after dividing the image into smaller areas. This code uses a thinning process to obtain the skeleton of the filaments based on the *HitOrMiss* transform (Sonka et al., 1999) that removes the branches of the skeleton after iteratively computing the end points.

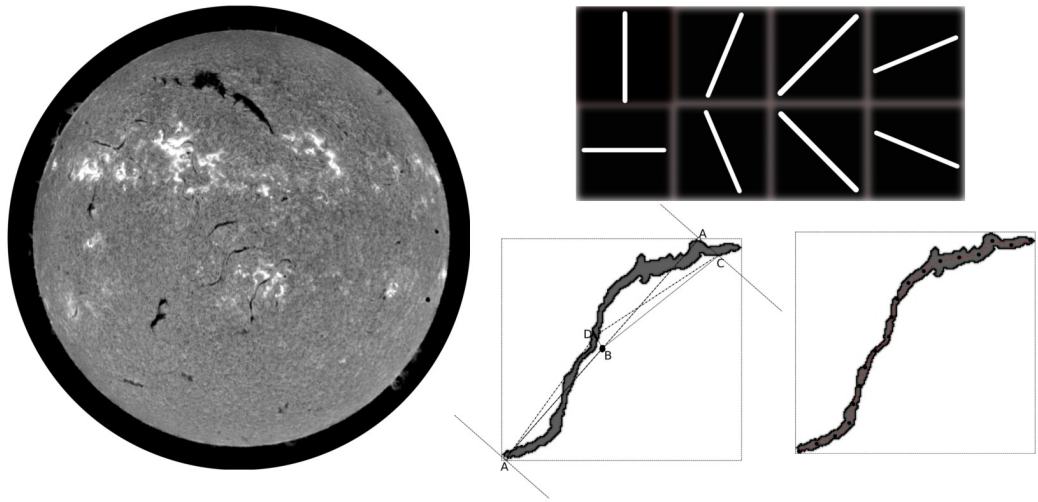

*Figure 4.* Full disk H-alpha image where the filaments can be seen as long dark and thin structures. On the right side; top: the eight directional linear structuring elements used by advanced morphological filter. On the bottom the first step and final result of the algorithm that determine the filament's spine. The labels refer to the order in which the points are found.

## DETECTION OF TEMPORAL FEATURES

The previous section focused on detecting features that can be localised within a single image. However, there are other forms of solar features whose detection is more complicated and requires temporal information. This section looks at two of these features - CMEs and coronal dimmings. Both may be easily identified when viewed in a sequence of images (i.e., a movie) but remain difficult to detect in a single image.

### Coronal Mass Ejections

CMEs are large-scale eruptions of plasma and magnetic field from the surface of the Sun. They travel through interplanetary space with velocities of up to several thousand kilometres per second, and have consequences for space-borne instruments and planetary atmospheres, often manifested as auroras on Earth and indeed other magnetic-field-protected planets (e.g., Saturn; Prangé et al., 2004). The diffuse and transient appearance of CMEs in images makes them difficult to automatically identify and track. They are best observed with the assistance of coronagraphs, a telescope attachment designed to block out the solar disk in order to better observe the surrounding corona (which is orders of magnitude fainter than the disk). They essentially create an artificial eclipse and are currently used in the Large Angle and Spectrometric Coronagraphs (LASCO; Brueckner et al., 1995) C2/3 instruments onboard *SoHO* and the Sun-Earth Connection Coronal and Heliospheric Imagers (SECCHI; Howard et al., 2008) COR1/2 instruments onboard the *Solar TErrestrial RElations Observatory* (*STEREO*; Kaiser et al., 2008). SECCHI also contains two wide-angle, visible-light imaging systems called the Heliospheric Imagers (HI).

A variety of catalogues exist that are maintained by individual instrument teams: [CDAW Catalog][3] and [NRL LASCO CME List][4] from *SoHO*, and the [COR1 CME Catalog][5] and [HI1 Event List][6] from *STEREO*. These catalogues provide information on the timing and properties of a CME, including the position angle, angular width, height, velocity, and acceleration. However, the creation and population of these catalogues are time consuming and the measured parameters are subject to human bias as they include manually performed processing. The automation of CME detection is highly desirable. To date, several automated CME detection algorithms have been proposed for use with LASCO C2 data and extensible to COR1. The Computer Aided CME Tracking (CACTus) algorithm was the first attempt at automation (Berghmans et al., 2002). This was followed by the Solar Eruptive Event Detection System (SEEDS; Olmedo et al., 2008) algorithm from George Mason University and the Automatic Recognition of Transient Events and Marseille Inventory from Synoptic maps (ARTEMIS; Boursier et al., 2009) from the LASCO team at Laboratoire d'Astrophysique de Marseille. All of these automated codes rely on the use of more than one frame to detect a CME. More recently, Byrne et al. (2009) propose a method to overcome this problem for real-time detection in single images.

The pre-processing of the images followed by these groups differs from the standard methods proposed by the instrument teams. This is because the standard reduction is not optimized to detect CMEs (e.g., the presence of background stars, planets, and comets is usual within these images). It is worth noting that LASCO is the most successful comet-finder in history, having detected over *one thousand six-hundred* comets in over thirteen years of operation[7]#. The images are exposure time normalized, corrected for cosmic rays, stars or planets by different methods, and transformed to their preferred coordinate system. Figure 5 shows an example of the different transformations used by each of the methods listed previously. CACTus transforms each from the native Cartesian coordinate system to a polar coordinate system [r, position angle], where r is the radial distance from the centre of the Sun and the position angle is the angle (anticlockwise) from a certain reference point (the ecliptic). The transformed images are stacked to produce a [r, position angle, t] data cube, which is iteratively processed to estimate the background and to remove the dust corona and rotating streamers. Following this, [r, t] slices are extracted from the cleaned datacube to proceed with the CME detection. ARTEMIS creates synoptic maps that consist of the generation of [position angle, t] images, complementary to the CACTus approach. Finally, SEEDS works in polar coordinates [r, position angle] after determination of the running-difference between two consecutive images. The method proposed in Byrne et al. (2009) does not require a coordinate transformation prior to the CME detection but, as with the other catalogues, it may be strengthened by utilising the temporal information across frames.

Figure 5 demonstrates that the same CME can presents a different signature in each of the transformations, therefore the techniques for the resulting detections are different. CMEs appear as inclined lines in CACTus, which relies on the Hough transform for detection. In ARTEMIS the CMEs appear with different morphologies, which are classified into four types: undistorted vertical streaks without temporal dispersion; quasi-symmetric arc shapes; arc shapes followed by a second structure with a dark zone in between; all remaining events with unclear signature. ARTEMIS detection involves three main steps of filtering, segmentation, and merging with high-level knowledge. Filtering is carried out line by line, removing the background with a median filter of 7-pixel width. The segmentation process is performed by a simple thresholding process with a value selected by *experience*, followed by the application of the Line Adjacency Graph (LAG; Pavlidis, 1986). This step removes small artificial "holes" by performing a morphological closure operation, identifies regions of interest, computes their geometrical and statistical parameters, and removes those that are smaller than a certain size. Finally, regions of interest are associated to the same CME if they simultaneously satisfy three empirically determined conditions that form the high-level knowledge.

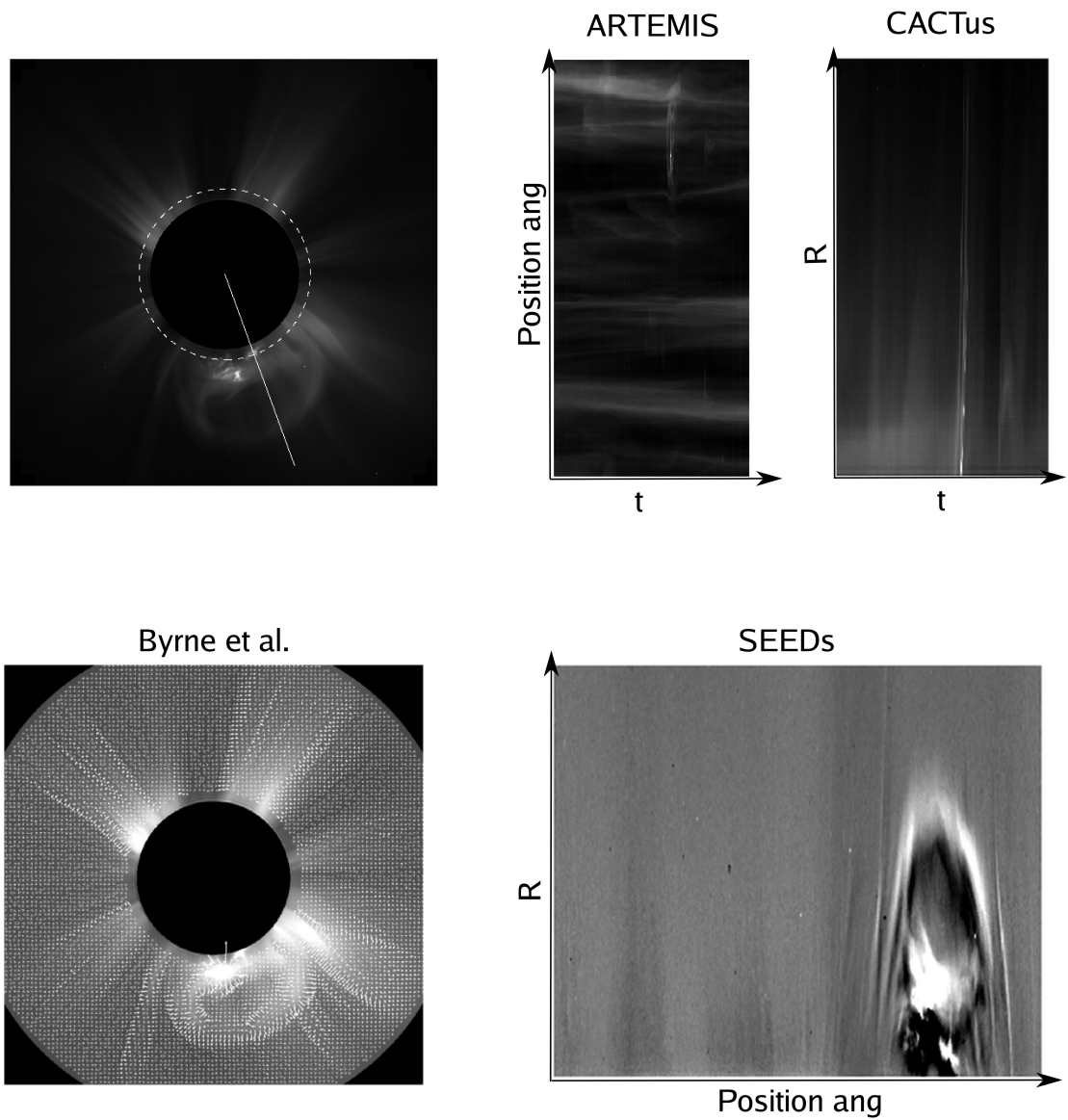

*Figure 5. Example of a CME as interpreted by the different algorithms presented here. The top left image shows a single frame of the CME as observed by SoHO/LASCO C2. The dashed line indicates the radial intensity profile as determined by ARTEMIS for multiple time steps, while the solid line indicates the fixed angle intensity profile as determined by CACTus for multiple time steps (top right images).The bottom left image illustrates the vector representation of the multiscale detection outlined in Byrne et al. (2009) whereby the curvilinear structure of the CME front is exploited. The bottom right image is a running-difference image as determined by SEEDS in order to threshold the CME intensity structure.*

A CME in SEEDs appears as a bright leading-edge enhancement (positive values) followed by a dark area deficient in brightness (negative values), with the background appearing grey (zero change). A running-difference process removes quasi-static features, such as coronal streamers. The CME is extracted by a threshold-segmentation technique and a region-growing algorithm. Positives values of the image are projected into one dimension along the angular axis, obtaining the angular intensity profile. The threshold value is obtained from this profile as a number of standard deviations (the number of standard deviations is chosen by experimental methods, and its value is often between two and four). This gives the angle of the core that is grown to cover the whole CME. The region growing algorithm connects those values between the maxima of the core-angle and a second threshold, calculated as before but just over the values outside the core-angle.

Byrne et al. (2009) apply the multiscale edge-detection method as proposed by Young & Gallagher (2008) for CME detection within a single image frame. While not presently implemented in a catalogue, the potential for automation is clear and currently under development. The *wavelet* algorithm is based upon a high and low pass filtering technique which serves to decompose the image into multiple scales. A particular scale of the decomposition best improves the signal-to-noise ratio of the CME against the background, making it easier to detect the CME front edges in a single image. The size and directional information of the filters used to decompose the image provide magnitude (i.e., edge strength) and angular information (i.e., edge normals) for the structures in the image. It becomes possible to represent the image data as a mesh of vectors across the frame (Fig. 5) by combining the magnitude and angular information. Thresholding the areas of maximal edge strength corresponding to large spreads in the angular information along the curved CME front distinguishes the CME from the linear streamers in the image. Furthermore, a pixel-chaining routine may be implemented to outline the CME front edges which are then characterised with, say, an ellipse fitting routine. This detection algorithm is further strengthened by considering more than one scale in the image decomposition, effectively combining the magnitude and angular information across all scales for which the CME signal-to-noise ratio remains high. If temporal information is available (as it is when backdating a catalogue) the method may be refined by turning the single image thresholding into a spatiotemporal filter that considers the movement of the CME edges through frames as an additional constraint on the CME front detection and characterisation. Gallagher et al. (2010) further exploits the curvilinear nature of CMEs through the use of curvelets as a multiscale tool. This approach may further enhance the CME structure on scales which neglect the linear structures of streamers in the image data, which would be of great benefit to an automated detection algorithm.

The second step for any CME catalogue is the characterisation of the CME kinematics. It is not as much connected with image analysis as it is with the application of a model. However, the nature of the Hough transform used by CACTus constrains the CMEs to have constant velocity. Boursier et al. (2009) show a comparison of the catalogues produced by CACTus, SEEDs, and ARTEMIS together with the man-made CDAW showing that the automated catalogues tend to report more than twice as many CMEs as are identified by visual detection. However, the primary interest for the development of the automated algorithms is not to reproduce the human-biased results, but instead to be able to produce robust statistical analyses.

## Coronal Dimmings and Bright Fronts

Coronal dimmings are usually observed as decreases in intensity in soft X-rays (Hudson et al., 1996) and EUV data (Thompson et al., 1998). The cause of these dimmings is still under debate, but the two most accepted possibilities are either a density depletion caused by an evacuation of plasma or a change in the bulk plasma temperature out of the passband of the image filter. The importance of these events and their physical cause is related to the potential of using these events to predict CMEs in the absence of coronograph images. This is currently of particular interest because NASA's recently launched *Solar Dynamics Observatory* (*SDO*) has no coronagraph instrument onboard.

The visualisation of these transient events is often optimised using base- or running- difference imaging. The base-difference method differs from the running difference method in that all images in the series have the same fixed image (usually obtained prior to the event) subtracted from them, whereas running-difference subtracts the previous image in time. Figure 6 shows a typical case of coronal dimming using running-difference imaging, where the central dark area (i.e., lower intensity than the previous image) in the middle panel expands outward over the solar surface in the right panel.

The Novel EIT wave Machine Observing (NEMO; Podladchikova & Berghmans, 2005) algorithm allows for real-time analysis of *SoHO*/EIT data. The algorithm works in two phases - detecting an event occurrence, and extracting the information of the dimming. The detection of event occurrence is performed through statistical analysis methods based on the histogram distribution of running-difference intensities. The start of an event is characterised by a sudden increase of variance, while the sign of skewness changes and the kurtosis increases rapidly during the event. Once the start and duration of the event is known the algorithm proceeds to the extraction of the dimming. Fixed difference images are then used to extract the dimming from the background. Pixels are collected into two groups - a maximal and minimal pixel map. The maximal pixel map comprises those pixels with values that fall below -1sigma on the histogram, where sigma is the value of the variance before the event occurs. Simultaneously, the minimal pixel map is constructed by selecting the darkest 1% of all pixels from the fixed-difference image. A median filter removes all of the smaller structures considered as "noise". The final dimming region is then extracted using the minimal pixel map as seeds for a region-growing method, keeping the condition of a simply-connected region and restricting it to pixels from the maximal pixel map. After a region is extracted, the area, location, volume, mass and light curves are obtained for each event. Recently, Attrill & Wills-Davey (2009) modified NEMO to adapt it to the Atmospheric Imaging Assembly, which is the successor of *SoHO*/EIT onboard *SDO*.

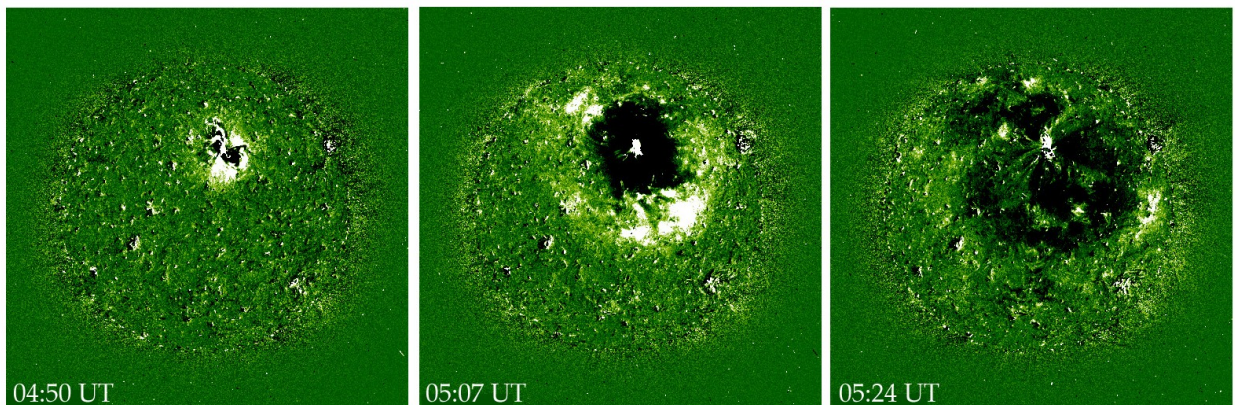

*Figure 6. Coronal dimming seen in 195 Å SoHO/EIT running difference images. Courtesy of SoHO/NASA website.*

**OUTLOOK**

Observatories from all around the world have been observing and classifying most of the features discussed here for decades. With the increase in data sizes in recent years, and due to the retirement of the experts on whom we relied to extract features manually, scientists have been turning their research toward the automation of such detections. The automation of solar feature detection gives a few advantages to the solar physicist. Firstly, and probably one of the most important advantages for statistical studies, is the creation of non-human biased catalogues. It is well known in observational science that when the observer is substituted, the records show a systematic variation. Secondly, the automation of the techniques accelerates the process of feature segmentation, which nowadays its crucial to manage the amount of data available.

The creation of a non human-biased catalogue is not an easy task. Solar features have been named after how they were observed for the first time, with their description in the hand-made catalogues depending on the observer. A comparison of how any automated algorithm performs, as compared with the previous catalogues, will always show differences. This is mainly due to the parameters used to define the feature itself. Therefore, improvements in detection are always tied with a comparison to the feature description.

The relatively sudden increase in the amount of solar data available is changing the way scientists work. There are two new space missions which will provide a huge amount of high-quality data. The first mission, launched in November 2009, is a European Space Agency mini-satellite called *Proba2*, which provides 1kx1k full-disk images of the 1 MK corona every minute and the ability to off-point from the solar disk (i.e., to track transient events such as CMEs). The second mission, launched in February 2010, is NASA's *SDO*. It will produce a 4kx4k full-disk image in 10 different filters every 10 seconds, and is also equipped with a high-resolution magnetometer that will far surpass the spatial and temporal resolution of *SoHO*/MDI. The analysis of this new data cannot be carried out in the traditional way, due to limitations in storage and bandwidth. The researcher will not be able to download a whole day of data (1.5 TB) to his personal computer. The new approach is to download only the desired feature, made possible by a pipeline of automated analysis algorithms working on each new image downloaded from the spacecraft which will feed a catalogue with feature properties. These catalogues realize the idea of virtual observatories (VO) as a collection of multiple data archives and tools to facilitate multi-instrument research. Recently the VO ideal has matured into a broader concept than simply a collection of data and tools. This is the case with the HELiophysics Integrated Observatory (HELIO[8]), a new project that provides possible links of events throughout the whole heliosphere in addition to data access.

These new instruments provide higher spatial and temporal resolution that could change feature detection either for better or worse. On the bright side, features may be better defined (i.e., boundaries will be more accurate). However, that could also make them more difficult to detect (i.e., previously contiguous bright areas may now be disconnected). This issue provides an advantage to the feature detection techniques based on fuzzy clustering (SPOCA) or supervised clustering (Dudok de Wit, 2006), where the algorithm segments the image based solely on the data provided. Relating these features to current classifications may be a difficult task. It seems clear that the definition of features cannot be linked to what is visible in only one wavelength (e.g., the detection of coronal holes can be differentiated from filaments by using the magnetic field information). More robust feature definitions will be achieved with the help of *SDO* by combining imaging of the chromosphere and corona in multiple passbands with photospheric vectormagnetograms, which provide the orientation and magnitude of the magnetic field at the solar surface.

The future perspectives for image processing applications to space weather are clear: near real-time feature extraction and analysis is crucial for forecasting. The solar physics community is well aware of this issue as is evident in the success of the Solar Image Processing Workshop series[9]. The detection of active regions, filaments, coronal holes, and tracking of CMEs is key, though a better understanding of these processes is required. An important advance would be the detection of transient events in single frames, which could, e.g., detect CMEs in single coronagraph images without the need for an image sequence. The biggest advancement could perhaps be in the search for pre-cursors for these events, and it seems that multiscale techniques may be a vital tool in this case.

[1] http://www.noaa.gov
[2] http://www.ngdc.noaa.gov/stp/SOLAR/ftpsunspotnumber.htm
[3] http://cdaw.gsfc.nasa.gov/CME_list/
[4] http://lasco-www.nrl.navy.mil/index.php?p=content/cmelist
[5] http://cor1.gsfc.nasa.gov/catalog/
[6] http://www.sstd.rl.ac.uk/stereo/HIEventList.html
[7] http://sungrazer.nrl.navy.mil/
[8] http://www.helio-vo.eu
[9] http://www.sipwork.org